\documentstyle[epsfig]{article}
= -2truecm
\begin{document}
\begin{center}

{\large\bf Exact boson mapping of the reduced BCS pairing
Hamiltonian} \vskip .6cm {\normalsize Feng Pan$^{a,~b}$ and J. P.
Draayer$^{b}$} \vskip .2cm {\small $^{a}$Department of Physics,
Liaoning Normal University, Dalian 116029, P. R. China\vskip .1cm
$^{b}$Department of Physics and Astronomy, Louisiana State
University, Baton Rouge, LA 70803-4001, USA}
\end{center}

\begin{abstract} An exact boson mapping of the reduced  BCS (equal
strength) pairing Hamiltonian is considered. In the mapping, fermion
pair operators are mapped exactly to the corresponding bosons. The
image of the mapping results in a Bose-Hubbard model with level
dependent hopping. Though the resultant Bose-Hubbard Hamiltonian is
non-Hermitian, all eigenvalues are real when $Uk/t< 1$, where $k$ is
the total number of bosons. When $U/t=1$, a part of spectrum of the
Bose-Hubbard Hamiltonian corresponds exactly to the whole spectrum
of the reduced BCS pairing Hamiltonian.
\end{abstract}

Pairing is one of important residue interactions in many areas of
physics, especially in the studies of superconductors,$^{[1,~2]}$
nuclear systems,$^{[3]}$ metallic clusters,$^{[4,~5]}$ and
liquids.$^{[6]}$ The limitations of the Bardeen-Cooper-Schrieffer
(BCS) and Hartree-Fock-Bogolyubov (HFB) methods$^{[1,7]}$ for
finding approximate solutions of finite nuclear systems and
nanoscale metallic grains are well understood.$^{[4,~8]}$
Fortunately, the reduced BCS (equal strength) pairing model was
proved to be exactly solvable following Richardson's early
work$^{[9,~10]}$ and studies based on the Gaudin algebraic Bethe
ansatz method,$^{[11]}$ which has received a lot of attention
recently.$^{[12,~13]}$

On the other hand, there is also a long history in search for
appropriate boson mapping methods or boson expansions for nuclear
many-body systems,$^{[14]}$  which can also be carried out for other
many-fermion systems. It is well known that in most circumstances
pairs of fermions exhibit boson-like behavior. In such approaches,
the degrees of freedom of fermion pairs are directly replaced by
exact boson degrees of freedom. These methods are potentially
helpful in describing collective motion in terms of boson degrees of
freedom to avoid usual difficult fermionic formulation since boson
operators have their counterparts in classical canonical variables,
and thus provide a direct link between microscopic nuclear models
and phenomenological collective models. A lot of attention has been
paid$^{[14]}$ especially after the success of the interacting boson
model for nuclei.$^{[15]}$

The purpose of this letter is to report an exact boson mapping of
the reduced BCS pairing Hamiltonian. When $m$ single fermions occupy
the $j_{1}$, $j_{2}$,$\cdots$, $j_{m}$ levels, respectively, the
reduced BCS pairing Hamiltonian for deformed nuclear system is given
by

$$\hat{H}_{\rm
BCS}/G=\sum_{\tau=1}^{m}(\epsilon_{j_{\tau}}/G)
(c^{\dagger}_{j_{\tau}\uparrow}c_{j_{\tau}\uparrow}+
c^{\dagger}_{j_{\tau}\downarrow}c_{j_{\tau}\downarrow})
+\sum_{j}~^{\prime}\eta_{j}\hat{k}_{j}-\sum_{i,j}~^{\prime}S^{+}_{i}
S^{-}_{j},\eqno(1)$$ where $c^{\dagger}_{j\sigma}$ ($c_{j\sigma}$)
are fermion creation (annihilation) operators,
$S^{+}_{j}=c^{\dagger}_{j\uparrow}c^{\dagger}_{j\downarrow}$ and
$S^{-}_{j}=c_{j\downarrow}c_{j\uparrow}$ are pair creation
(annihilation) operators,
$\hat{k}_{j}=(c^{\dagger}_{j\uparrow}c_{j\uparrow}+
c^{\dagger}_{j\downarrow}c_{j\downarrow})/2$, $\epsilon_{j}$ are
single-particle energies taken from any deformed mean-field theory,
$G>0$ is the equal strength pairing parameter,
$\eta_{j}=2\epsilon_{j}/G$, and the summation sign with prime
indicates that the sum is restricted to levels other than those
occupied by the single fermions.

Because solutions of $m\neq 0$ cases are basically similar to those
of seniority zero case, in the following we only consider the case
with $m=0$. For $k$-particle excitation, the wavefunction of (1) in
this case with $p$ levels considered can be written as$^{[9,~10]}$

$$\vert
k;\xi\rangle=S^{+}(E_{1}^{(\xi)})S^{+}(E_{2}^{(\xi)})\cdots
S^{+}(E_{k}^{(\xi)})\vert 0\rangle,\eqno(2)$$ where $\vert 0\rangle$
is the pairing vacuum state satisfying $S^{+}_{j}\vert 0\rangle=0$
for $1\leq j\leq p$,

$$S^{+}(E_{\mu}^{(\xi)})=\sum_{j=1}^{p}
{1\over{\eta_{j}-E_{\mu}^{(\xi)}}}S^{+}_{j}, \eqno(3)$$ with the
corresponing eigen-energy
$E^{(\xi)}_{k}=G\sum_{\mu=1}^{k}E_{\mu}^{(\xi)}$.

The pair energies $E_{\mu}^{(\xi)}$ should satisfy $k$ coupled Bethe
ansatz or Richardson-Gaudin equations:

$$1=\sum_{j=1}^{p}{1\over{\eta_{j}-E_{\mu}^{(\xi)}}}+\sum_{\nu\neq\mu}
{2\over{E_{\mu}^{(\xi)}-E_{\nu}^{(\xi)}}}\eqno(4)$$ for
$\mu=1,2,\cdots,k$. It is understood that the additional quantum
number $\xi$ in (2)-(4) is introduced to label the $\xi$-th set of
roots $\{E_{\mu}^{(\xi)}\}$ of the equations (4).

To map the reduced BCS Hamiltonian (1) into a boson Hamiltonian, we
first use the mapping that maps the fermion pair operators
$\hat{k}_{j}$, $S^{\pm}_{j}$ into the corresponding real boson
operators with

$$\hat{k}_{j}\mapsto n_{j}=b^{\dagger}_{j}b_{j},~
S^{+}_{j}\mapsto b^{\dagger}_{j},~~S^{-}_{j}\mapsto
b_{j}~~\forall~j,\eqno(5)$$ in which the images satisfy the usual
commutation relations of boson operators:
$[b_{i},b^{\dagger}_{j}]=\delta_{ij}$, and $[b_{i}, b_{j}]=0$. It is
clear that this mapping is different from that based on group
structure$^{[14]}$ because the images of $S^{\pm}_{j}$ no longer
satisfy the commutation relations of the original SU(2) algebras.
Furthermore, the mapping is unitary and number-conserving. We then
seek a Bose Hamiltonian constructed from those boson images which
should keep the wavefunction (3) consistent after the mapping. We
found the one-body term in (1) does keep the same form after the
mapping, while the pairing interaction term can not be mapped into
one-body form, but with an additional non-Hermitian two-body
interaction term, which is quite natural because the fermion pairing
interaction like hard-core boson hopping can not be replaced by
usual boson hopping. The final image of (1) after the mapping (5) is
of the following form:

$$\hat{H}_{\rm Bose}/G=\sum_{j=1}^{p}(\eta_{j}-1)n_{j}-
\sum_{i\neq j}b^{\dagger}_{i}b_{j}+
\sum_{i,j=1}^{p}n_{j}b^{\dagger}_{i}b_{j}.\eqno(6)$$

To reveal the dynamics of the Hamiltonian (6), let us consider a
more general form of (6) with

$$\hat{H}_{\rm BH}=\sum_{j}(2\epsilon_{j}-t-U)n_{j}-
\sum_{i\neq j}(t-n_{j}U)b^{\dagger}_{i}b_{j}+
U\sum_{j}n_{j}^{2},\eqno(7)$$ where $2\epsilon_{j}-t-U$ in the first
term can be regarded as contribution from external potential or
on-site disorder, the second term describes boson hopping among all
sites with site dependent hopping parameter $t-n_{j}U$, and the
third term is the on-site repulsion. Since the two-body interaction
term usually contribute with the same order of magnitude as the
one-body term, the on-site repulsion parameter may be set as
$U=U_{0}/k$, where $k$ is the total number of bosons, in which
$U_{0}$ and $t$ is of the same order of magnitude. Hence, the
reduced BCS pairing Hamiltonian is mapped into a Bose-Hubbard model
with site-dependent hopping parameter $t-U_{0}(n_{j}/k)$. Therefore,
the more the bosons on the $j$-th level, the less the hopping
strength of other bosons hopping onto $j$-th level if $t\geq U_{0}$.
In the Bose Hamiltonian (6) with $t=U=G$, however, the condition
$1\geq n_{j}$ is no longer satisfied if $n_{j}\neq 0$ or $1$, which
means that the fermion pairing interaction looks extremely repulsive
after the boson mapping (5).

To prove that (6) is indeed the exact boson image of (1), one can
simply verify that the wavefunctions of (7), at least a part of
them, can indeed be written as the boson image of (2) with

$$\vert k,\xi\rangle=B^{+}(E_{1}^{(\xi)})B^{+}(E_{2}^{(\xi)})\cdots
B^{+}(E_{k}^{(\xi)})\vert 0\rangle_{\rm B},\eqno(8)$$ where $\vert
0\rangle_{\rm B}$ is the corresponding boson vacuum, and

$$B^{+}(E_{\mu}^{(\xi)})=\sum_{j=1}^{p}
{1\over{2\epsilon_{j}/t}-E_{\mu}^{(\xi)}}b^{\dagger}_{j}, \eqno(9)$$
with the corresponding eigen-energies

$$E^{(\xi)}=t\sum_{\mu=1}^{k}E_{\mu}^{(\xi)}\eqno(10)$$
and the Bethe ansatz equations

$$1=\sum_{j=1}^{p}{1\over{2\epsilon_{j}/t-E_{\mu}^{(\xi)}}}+\sum_{\nu\neq\mu}
{2(U/t)\over{E_{\mu}^{(\xi)}-E_{\nu}^{(\xi)}}}\eqno(11)$$ for
$\mu=1,2,\cdots,k$.

Though it is difficult to analyze the spectrum generated by (11)
analytically, one can verify that the whole spectrum described by
(10) obtained from solutions of (11) are real and complete with $(k
+ p - 1)!/(k!(p - 1)!)$ eigenvalues when $Uk/t\leq 1$. Simple
examples for $p=3$,  $k=2$, $3$ are shown in Table 1.

\vskip .3cm \noindent{\bf Table 1.}~{The pair energies or roots of
(11) and eigen-energies given by (10) for $p=3$ and $k=2$ and $3$
with $t=1$, $\epsilon_{1}=1.1$, $\epsilon_{2}=2.2$,
$\epsilon_{3}=3.3$,
and $U=1/3$. The dimension is exactly equal to $(k+p-1)!/((p-1)!k!)$.}\\
\begin{tabular*}{\textwidth}{ccccccc}
\hline \hline $k$ &Dimension~&Eigenvalues &~~~Roots\\
\hline\\
2&6&1.768&~~~~~~$E_{1}=0.884-0.792i$&$E_{2}=0.884+0.792i$\\
&&4.475&~~~~~~$E_{1}=0.935$~~~~~~~~~~~~&$E_{2}=3.540~~~~~~~~~~~~$ \\
&&6.763&~~~~~~$E_{1}=0.735$~~~~~~~~~~~~&$E_{2}=6.028$~~~~~~~~~~~~\\
&&7.622&~~~~~~$E_{1}=3.811-0.358i$&$E_{2}=3.811+0.358i$\\
&&9.744&~~~~~~$E_{1}=3.770$~~~~~~~~~~~~&$E_{2}=5.974$~~~~~~~~~~~~ \\
&&12.428&~~~~~~$E_{1}=6.214-0.258i$&$E_{2}=6.214+0.258i$\\\\

3&10&3.930&$E_{1}=1.030$&$E_{2}=1.450- 1.229i$&$E_{3}=1.450+ 1.229i$\\
&&6.097&$E_{1}=3.343$&$E_{2}=1.377-0.589i$&$E_{3}=1.377+0.589i$\\
&&8.237&$E_{1}=5.985$&$E_{2}=1.126-0.685i$&$E_{3}=1.126+0.685i$\\
&&8.577&$E_{1}=1.227$&$E_{2}=3.675-0.425i$&$E_{3}=3.675+0.425i$\\
&&10.704&$E_{1}=3.672$&$E_{2}=1.104$~~~~~~~~~~~~&$E_{3}=5.928$~~~~~~~~~~~~\\
&&12.108&$E_{1}=3.896$&$E_{2}=4.106-0.568i$&$E_{3}=4.106+0.568i$\\
&&13.302&$E_{1}=0.932$&$E_{2}=6.185-0.276i$&$E_{3}=6.185+0.276i$\\
&&13.718&$E_{1}=5.822$&$E_{2}=3.948-0.302i$&$E_{3}=3.948+0.302i$\\
&&16.142&$E_{1}=3.848$&$E_{2}=6.147-0.304i$&$E_{3}=6.147+0.304i$\\
&&19.184&$E_{1}=6.286$&$E_{2}=6.449-0.357i$&$E_{3}=6.449-0.357i$\\
\hline \hline\\
\end{tabular*}

When $U=t=G$, with which (7) is reduced to (6) corresponding to the
image of the reduced BCS pairing Hamiltonian (1). In this case,
though it is not guaranteed that all eigenvalues of (7) are real,
especially for large $k$ cases, a part of them satisfying (11)
consisting of $p!/((p-k)!k!)$ eigenvalues, which are the same as
those given by (4), correspond exactly to those of the reduced BCS
pairing Hamiltonian (1). There are
[$(k+p-1)!/((p-1)!k!)-p!/((p-k)!k!)$] more other eigenvalues which
are not provided by (10)-(11). Hence, eigenvectors other than those
corresponding to the eigenvalues given by (10) can not be written in
the Bethe ansatz form (8).

Similar to the Dyson mapping$^{[14]}$ and, for example, the
iterative boson expansion approach,$^{[16]}$ the resultant Bose
Hamiltonian (6) is non-Hermitian. In addition, there will be
spurious states involved in the boson space. Therefore, the Bose
Hamiltonian should be projected onto the physical subspace. Let
$\hat{P}$ be the projection operator, which can be expressed as

$$\hat{P}=\sum_{k\xi}\vert k,\xi)(\xi,k\vert,\eqno(12)$$
where $\vert k,\xi)$ are normalized eigenvectors given by (8), and
the sum runs over all possible values according to the number of
solutions of (11) with $U=t=G$. Thus, it is clear that the
projection operator $\hat{P}$ annihilates unphysical
subspace.$^{[14]}$ It follows that the projected Bose Hamiltonian
with

$$\tilde{H}_{\rm Bose}=\hat{P}\hat{H}_{\rm Bose}\hat{P}\eqno(13)$$
is digonalizable under the physical subspace spanned by $\{\vert
k,\xi)\}$ with results shown by (8)-(11). Hence, we obtain the Bose
Hamiltonian $\tilde{H}_{\rm Bose}$ exactly equivalent to the reduced
BCS Hamiltonian (1) in the physical boson subspace.

\vskip .4cm In summary, an exact boson mapping of the reduced  BCS
pairing Hamiltonian is obtained under the guidance of the
Richardson-Gaudin exact solutions of the reduced  BCS pairing model.
Though non-Hermitian, all solutions of the resultant Bose-Hubbard
Hamiltonian are real and provided by the Richardson-Gaudin type
wavefunctions and the corresponding Bethe ansatz equations when
$Uk/t< 1$. The physical Bose image of the reduced  BCS pairing
Hamiltonian is obtained by the projection method which is exactly
equivalent to the reduced BCS Hamiltonian (1) in the physical boson
subspace. Because what we have studied is based on deformed shell
model like basis, the boson operators $\{b_{j},b^{\dagger}_{j}\}$ do
not conserve angular momentum which must be restored by angular
momentum projection as mentioned in [17]. In the Nilsson mean-filed,
for example, the boson operator $b_{i}^{\dagger}$ can be rewritten
in terms of spherical pairs as

$$b_{i}^{\dagger}=x^{i}_{0}s^{+}_{0}+x^{i}_{2}d^{\dagger}_{0}
+x^{i}_{4}g^{\dagger}_{0}+\cdots,\eqno(14)$$ where $x^{i}_{L}$ are
transformation coefficients between the $i$-th Nilsson level and the
spherical basis, and $s^{+}_{0}$, $d^{\dagger}_{0}$, etc. are boson
operators with $L=0,2,\cdots$, and $M_{L}=0$. After the angular
momentum projection, one can better understand the intimate links
between mean-field plus pairing models and the interacting boson
model. Further study about this problem will be carried out in the
near future.

\vskip .5cm  Support from the U.S. National Science Foundation
(0500291), the Southeastern Universities Research Association, the
Natural Science Foundation of China (10575047), and the LSU--LNNU
joint research program (9961) is acknowledged.

\end{document}